# Diode Laser Welding of ABS: Experiments and Process Modelling

M.ILIE[1,3], E.CICALA[1,2], D.GREVEY[2], S. MATTEI[2], V. STOICA[1]

[1] Mechanical Engineering Faculty, "Politehnica" University of Timisoara,

Mihai Viteazu nr.1, Timisoara, 300222, Romania

[2] Institut Carnot de Bourgogne, UMR 5209 CNRS-Université de Bourgogne,

12 Rue de la Fonderie,71200 Le Creusot, France

[3] ISIM - National R&D Institute for Welding and Material Testing,

30, Mihai Viteazul Bv, 300222 Timisoara, Romania

**Abstract**

*In the present paper, the laser beam weldability of ABS plates is determined by combining experimental and theoretical aspects. In modelling the process, an optical model is used to determine how the laser beam is attenuated by the first material and to obtain the laser beam profile at the interface. Using this information as input data in a thermal model, the evolution of the temperature field within the two components can be estimated. The thermal model is based on first principles of heat transfer and utilizes temperature variation laws of material properties. Corroborating the numerical results with the experimental ones, some important insights concerning the fundamental phenomena that govern the process could be extracted. This approach proved to be an efficient tool in determining the weldability of polimeric materials and assures a significant reduction of time and costs with the experimental exploration, this one being made only for finding the optimal values.*

Corresponding author. Tel.: +40 256 200 222; fax: +40 256 492 797.

E-mail addresses: mariana.ilie@mec.upt.ro, milie@isim.ro

**Introduction**

In through-transmission laser welding of polimeric materials a focused laser beam is directed at two overlapping thermoplastic parts: first designed to be transparent to the laser wavelength and a second one, absorbent in IR spectrum. Between the laser beam and the components to be assembled there is a relative motion which allows a continuously operating laser beam to irradiate a line of a specific width. The heat generated by absorption is transmitted to the transparent component by conduction. The bonding between the two components occurs by the interpenetration of the molecular chains in this area. Since this phenomenon is very active in a "fluid" state of matter, the temperature at interface has to be between the temperature of solid-liquid transition and the initial temperature of degradation of the thermoplastic materials [1]. Comparing to the traditional welding techniques, the laser welding efficiency is strongly dependent on the materials properties. In consequence, obtaining a high-quality weld joint is conditioned by a good understanding of the material behavior under laser irradiation, based on a clear identification and modeling of the optical and thermal phenomena involved.

Concerning the optical phenomena, particularly the laser beam attenuation in polymers, there are some experimental approaches for quantifying the influence of fillers, pigments, reinforcement materials, the combined effect of thickness and color and so on [2], but for modeling this attenuation all the researchers are using the Beer-Lambert law [3, 4, 5] which is appropriate in the case of absorbent polymers but not for low absorbent and scattering media. The scattering of the laser beam is an important phenomenon because it has a great influence on the energy distribution within the materials, especially at the interface where the heat source has to be defined. An experimental solution to this necessity was given by Potente and Becker [6, 4] which measured the laser beam profile before and after the transparent component thus obtaining both the attenuation and the energy distribution at the interface. Within the present study the scattering of the laser beam in semitransparent low absorbent polymers is modeled combining Mie theory and the Monte-Carlo method. This numerical model allows a good estimation of laser beam profile at the polymers interface and therefore a well defined heat source.

In computing the temperature field within the materials and at the interface the heat conduction equation is used [3, 4, 5 ,6 ,7] taking in account a perfect contact between the two components subjected to welding. An estimation of the temperature at the interface is required for establishing the welding process parameters like laser power and welding speed. There are also studies that consider a contact conduction between the joining components [3] based on the roughness of the surface, the contact pressure and the elastic modulus of the materials, parameters that varies with temperature during the welding process. In the present study, a hypothesis of a perfect contact between the two components was considered due to a reduce thickness of the T leg (2mm) and the good quality of the contact surface.

Within the frame of the present study, experimental and theoretical aspects of the laser welding are combined in order to determine the weldability of natural and black pigmented ABS plates.

**Experimental set-up and materials**

The ABS (acrylonitrile/butadiene/styrene) is a complex mixture consisting of styrene-acrylonitrile copolymer, a graft polymer of styrene-acrylonitrile and polybutadiene and some unchanged polybutadiene rubber. The presence of these rubber occlusions makes him inhomogeneous from the optical point of view. Experimental investigations involved two types of ABS (acrylonitrile/butadiene/styrene) plates. The first component is a natural ABS, semitransparent to the laser wavelength as it is mentioned above, and the second one is a black pigmented ABS, opaque in near-infrared spectrum. The general optical and thermal characteristics for the two materials are listed in table 1. The values are given at ambient temperature ($\approx 24°C$). For the 2mm thickness ABS natural plates the transmission coefficient presented in table 1 indicates a strong scattering of the laser beam with negligible absorption. This conclusion was drawn after infrared measurements of the temperature raise during its laser irradiation.

The experiments were carried out using a diode laser LASERLINE with a maximum output power of 100 W and a focal length of 0.150 m. The laser wavelength is 0.940 $\mu m$ and the focal spot has a rectangular shape ($l_x$=1.7 mm, $l_y$=1.3 mm).

To improve the contact between the two polymers a specially designed clamping device has been used. The analyzed welding configuration was a T-type geometry for through-transmission laser welding. The experimental setup is shown in figure 1.

The weld failure force was determined with a MTS machine presented in figure 2 with a special fixture for the T-geometry. The traction speed was 5mm/min.

**Basics of the process modelling**

Process modelling of laser through-transmission welding of polimeric materials involves a good knowledge of the optical and thermal behaviour of the involved materials. In modeling the optical behavior of the laser beam through a polymer, an extinction phenomenon of the laser radiation has to be considered in which absorption and scattering effects can equally coexist or prevail to each other. The thermal aspects are solved by using the fundamental laws of heat transfer.

*Optical model*

When the first material is semitransparent to the laser wavelength (due to a crystalline structure or to the presence of pigments within their matrix), in establishing the energetic and geometrical characteristics of the heat source at the interface, a quantification of the laser beam scattering is necessary. Since the use of classical approximation based on the Beer-Lambert law can lead to inaccurate results in the case of scattering media, a new approach is used. Its detail description is given in previous works [8] [9]. Basically, it involves two steps: (a) based on measurements of the transmitted intensity (in % from the incident one) for three thickness of the same semitransparent material, the scattering characteristics of the media (particles diameter, the relative refractive index and the particle concentration) are computed by applying an inverse algorithm. This allows obtaining an equivalent material which produces the same attenuation and broadening of the laser beam as the real one. In the second step, the laser beam profile within the semi-transparent material is determined by using a numerical model based on Mie theory and Monte Carlo method.

For the ABS plates, the measurements of the transmitted intensity (table 2) indicate an equivalent material having the following scattering characteristics: particle diameter $d=0.46\mu m$, relative refractive index $n_{particle}/n_{matrix}=1.42$ and particle concentration $c=0.44\%$. Using these characteristics as input data in the numerical model (Mie +Monte Carlo), the transmitted laser beam profile was obtained. In spite the fact that the incident laser beam has a rectangular shaped profile with a uniform distribution of the energy (figure 3-a) the transmitted beam has an elliptical spot and a Gaussian energy distribution (figure 3-b,c), the polymer acting like an optical filter. The characteristic radii (the radius where the laser beam intensity decreases with $1/e^2$) for the elliptical spot were $r_x$=2.5 mm and $r_y$=2.69 mm.

*Thermal model*

In defining the heat source, the transmitted laser beam profile obtained from the optical model is used. In the second material, one considers a strong absorption of the laser radiation, according to Beer-Lambert law, and a heat transfer through conduction between the two parts. The heat source in the case of a couple semitransparent-absorbent polymers is described by:

$$q_v(x,y,z) = \begin{cases} 0 & z < d_1 \\ p_t(x,y) \cdot \alpha \cdot e^{-\alpha \cdot (z-d_1)} & z > d_1 \end{cases} \quad (1)$$

where $d_1$ is the thickness of the semitransparent polymer, $\alpha$ absorption coefficient of the second material and $p_t(x,y)$ is the power density for the laser beam at the interface. The latter one is calculated by taking in account the on axis power density for the incident beam and the ratio of the equivalent intensities ($I_i$, $I_t$), correlated with the number of "launched" and "received" photons and their declared energies:

$$p_t(x,y) = \frac{P}{l_x l_y} \cdot \frac{I_t}{I_i} \cdot e^{-2 \cdot \left(\frac{x^2}{r_x^2} + \frac{y^2}{r_y^2}\right)} \quad (2)$$

The evolution of the temperature field in the two components to be assembled is determined by the resolution of the general heat conductivity equation:

$$\vec{\nabla}\left[\lambda(T)\vec{\nabla}T\right] + q_v = \rho \cdot c(T) \cdot \frac{\delta T}{\delta t} \quad (3)$$

where λ(T) is the thermal conductivity, ρ the density, and c(T) the specific heat capacity.

The variation laws of the thermal properties λ(T) and c(T) with temperature are taken from literature [15].

For the heat transfer between the two elements and the surrounding medium the boundary condition can be written as:

$$-k(T) \cdot \vec{\nabla} T \cdot \vec{n} = h(T)(T_\infty - T) + \varepsilon\sigma\left(T_{amb}^4 - T^4\right) \quad (4)$$

where $\vec{n}$ is the normal vector of the surface, h- the heat transfer coefficient, $T_{amb}$- surrounding medium temperature, $T_\infty$- external temperature, $\varepsilon$- material emissivity, $\sigma$- black body emission coefficient.

The above model is employed in a finite element based code (COMSOL). For shortening the computing time, a symmetry plane is considered (figure 4). The primary output from the mathematical model of laser welding is the temperature distribution in the weld zone as a function of time.

*Preliminary tests*

For the preliminary experimental tests, the selected factors of influence were: laser power, P [W], welding speed, v [m/min] and clamping pressure, Press. [MPa]. The linear energy is calculated according to:

$$E_{lin} = (P \cdot 60)/(1000 \cdot v) \text{ [J/mm]} \quad (5)$$

To highlight the influence of the process parameters an experimental design was conceived and carried out [10], [11], [12]. The structure of the experimental design is presented in the table 3. Following welding, for each parametric combination the samples are assessed by two different means: a visual inspection used to monitor the consistency of the welds and look for evidence of thermal decomposition and a destructive test to determine the force required to fail the weld. As a representative value for the destructive test, the average maximum failure force F of four replicas was used.

Analyzing the data from the welding experiment involving a variation of the clamping pressure and laser power and taking in account the numerical simulation of the temperature field evolution in the considered system some important insights can be depicted:

- following the microscopic observations of the welded zone and considering the obtained values for the average failure force in the explored field, it has been established that between the two plates an adhesive joining, not a welded one was obtained;
- this situation is due to an important increase of the temperature at the interface above the processing temperature of the ABS [14]. Considering the high fluidity of the material at these temperatures and the clamping pressure applied, it is obvious that the material is ejected from the interfacial zone thus diminishing the interpenetration of the molecular chains in this area.

The evolution of the failure force as a function of *laser power – clamping pressure* (figure 5-a) indicates that the optimal zone has to be searched towards low laser power and high pressure.

The response surface for the simulated temperature at the interface (figure 5-b) shows an important influence of the laser power on the temperature evolution and consequently on the weld strength.

In order to estimate the process parameters that will allow obtaining an adequate temperature at the interface, a $3^2$ factorial design was carried out involving only numerical simulations (table 4). Since the desired temperature at the interface should have lower values than the processing one (the material temperature during the injection), the proper combination of process parameter can be taken from the response surface presented in figure 6. A rough estimation indicates a laser power around 16W and a laser speed about 0.6 m/min.

*Exploratory experiments*

Exploratory experimental designs used to advance towards the optimum [11], [13] were conceived based on the conclusions of the preliminary tests and by taking as starting point the best results of the preliminary tests. The structure of the experimental designs is detailed in table 5.

The qualitative analyses of the test plates showed that for the laser power between 14 and 17 W a high value of the failure force is recorded indicating a good quality of the welded joint in this domain. The simulated temperature at the interface in this interval is situated between 170 and 185°C (figure 7-a, b). For a laser power outside of this interval the failure force decreases due to an incomplete welding or an adhesive joining of the plates.

An additional experimental design was carried out to highlight the influence of the number of the laser beam passages on the resistance of the welded plates. It was noted that, usually, for the same parametric conditions a certain improvement of failure force is recorded if one carries out two passages (figure 8-a) with the laser beam compared to only one passage (figure 8-b).

One can appreciate that the optimal weld strength of the welded plates, for the studied configuration, are obtained for the following process parameters:

- laser power P = 15W;
- speed of welding v [ 0.5; 0.55] m/min;
- clamping pressure Pr = 0.4MPa;
- a number of 2 passages of the laser beam.

**Conclusions**

The study carried out showed the influences of the operational parameters on the weldability and the mechanical resistance of the assembled ABS plates. The adopted experimental strategy, based on the method of the experimental designs, made it possible to highlight in an effective way the fields of weldability. One could note that the laser power used is the most important parameter on the mechanical resistance and the weldability is assured in a narrow interval of power laser. An important clamping pressure and two laser passages are useful to improve the performances of the weld strength.

The correlation between the experimental and simulated results gives the possibility to understand some of the phenomena that take place during the polymers laser welding. In the same time, the proposed numerical model proved its capabilities, the indicated parameters domains being validated by the experimental results. Thus the model is an efficient tool in determining the weldability of polimeric materials contributing to a significant reduction of time and costs with the experimental exploration, this one being made only for finding the optimal values.

**Acknowledgements**

This paper is part of the project no. 71-088 – POLYWELDSYS in the frame of PNCDI II, Romania

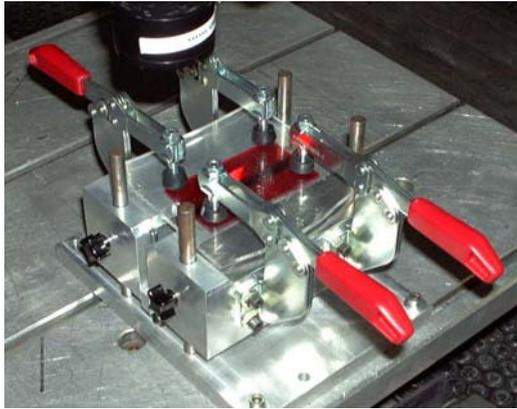
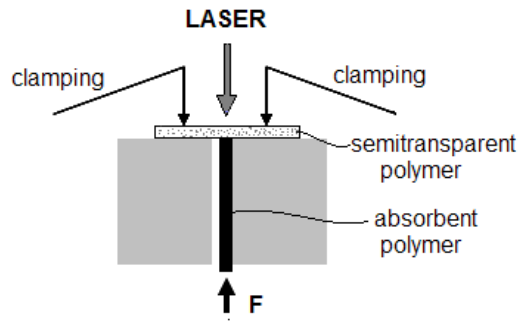

*Figure 1. Experimental set-up for laser welding of ABS plates*

*Table 1. Geometrical, optical and thermal properties of ABS plates (\* the values are given for λ=0.940 μm)*

| Properties | Dimensions | Horizontal plate<br>ABS natural | Vertical plate<br>Opaque ABS black |
|---|---|---|---|
| Thickness | $d$ [m] | $2.10^{-3}$ | $2.10^{-3}$ |
| Reflection factor* | $R$ [%] | 2.6 | 1 |
| Transmission factor* | $T$ [%] | 51.4 | - |
| Density | $\rho$ [kg.m$^{-3}$] | 1040 | 1010 |
| Heat capacity | $c_p$ [J.kg$^{-1}$C$^{-1}$] | 1350 | 1330 |
| Thermal conductivity | $\lambda$ [W.m$^{-1}$C$^{-1}$] | 0.19 | 0.2 |
| Glass transition temperature | $T_g$ [°C] | 114 | 120 |
| Processing temperature | $T$ [°C] | 221 | 225 |

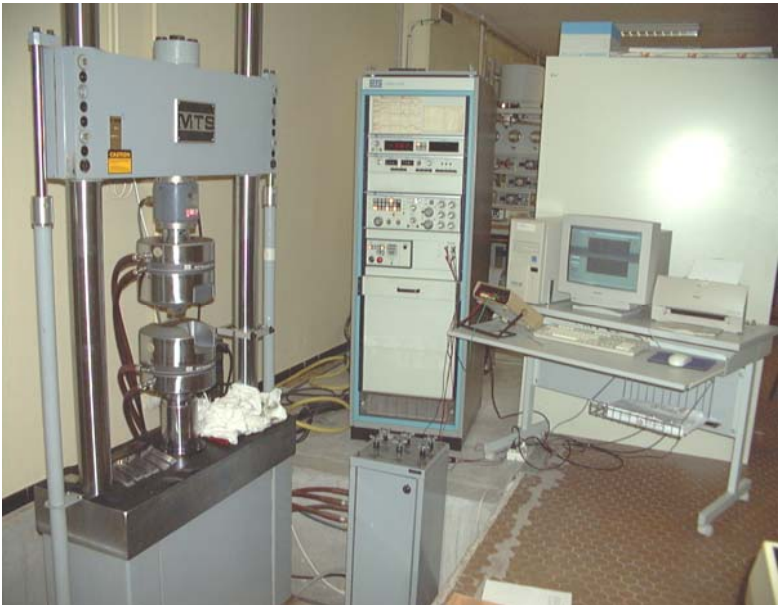
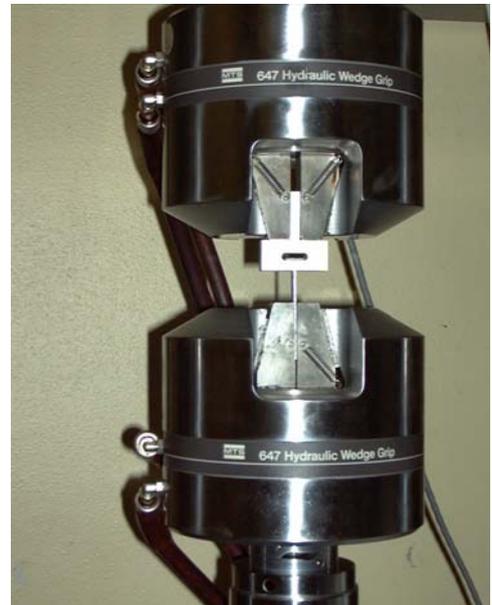

*a. photo of the experimental test bench*  *b. the probe fixing head*
*Figure 2 The experimental test bench for measuring the weld failure force*

Table 2 Transmitted intensity (in % from the incident one) for three thickness of the natural ABS plates

| Slab thickness [mm] | Measured $I_t/I_i$ [%] | Simulated $I_t/I_i$ [%] |
|---|---|---|
| 1 | 62.1 | 61.64 |
| 2 | 51.4 | 50.69 |
| 3 | 36.7 | 38.8 |

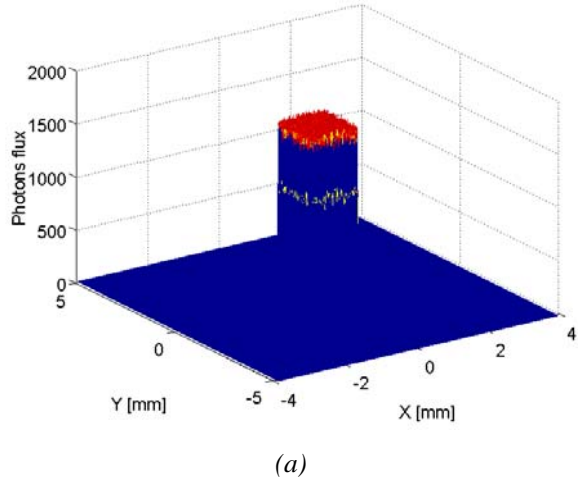
*(a)*

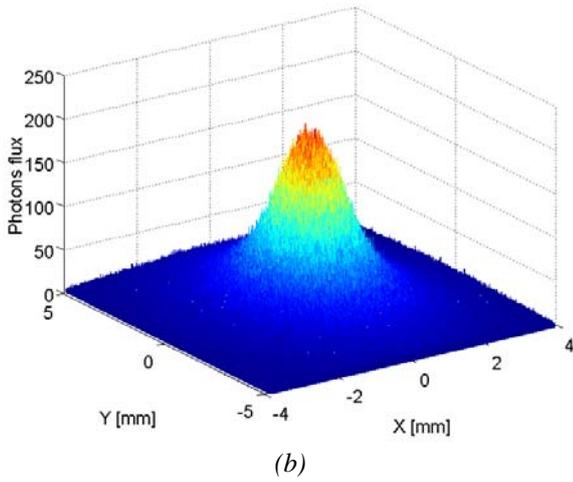
*(b)*

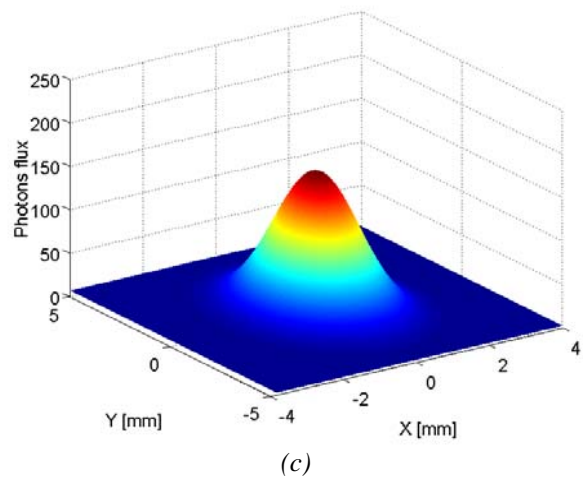
*(c)*

*Figure 3: The simulated profile of laser intensity distribution for the incident (a), transmitted through the 2 mm ABS slab (b) and the Gaussian approximation of the latter one (c)*

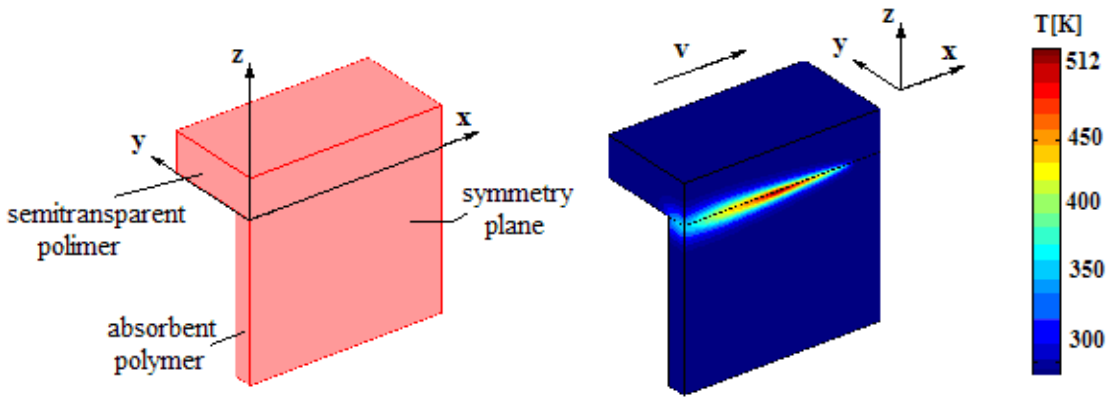

*Figure 4 T joint geometry considered for the numerical simulations*

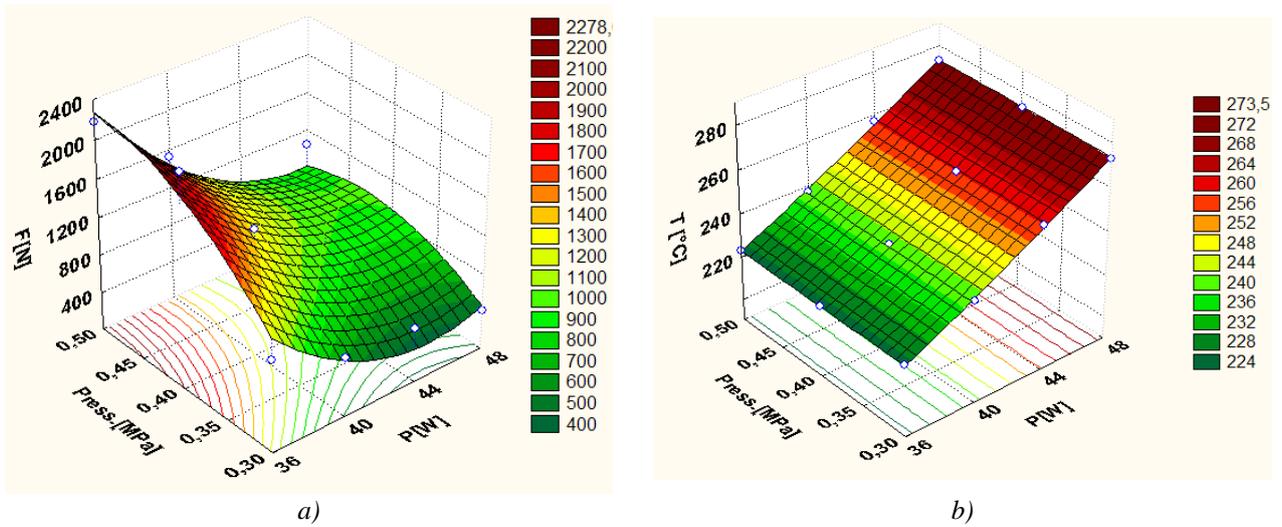

*a)* *b)*

*Figure 5 Response surface for the mean failure force (a) and the simulated temperature at the interface (b) in the explored domain*

*Table 4 $3^2$ factorial design used for numerical simulation of the temperature evolution*

| P[W] | v[m/min] | T[°C] |
|------|----------|-------|
| 10   | 0,3      | 173   |
| 30   | 0,3      | 390   |
| 10   | 0,65     | 121   |
| 20   | 1        | 157   |
| 20   | 0,3      | 287   |
| 30   | 0,65     | 267   |
| 30   | 1        | 212   |
| 20   | 0,65     | 200   |
| 10   | 1        | 90    |

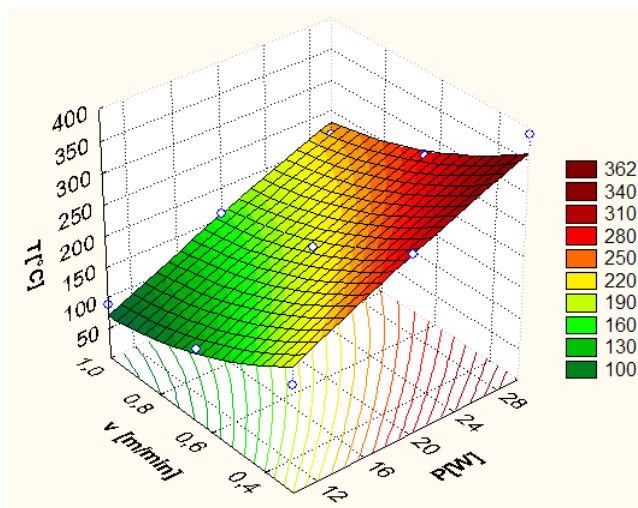

*Figure 6. Response surface for simulated temperature at the interface in the explored domain*

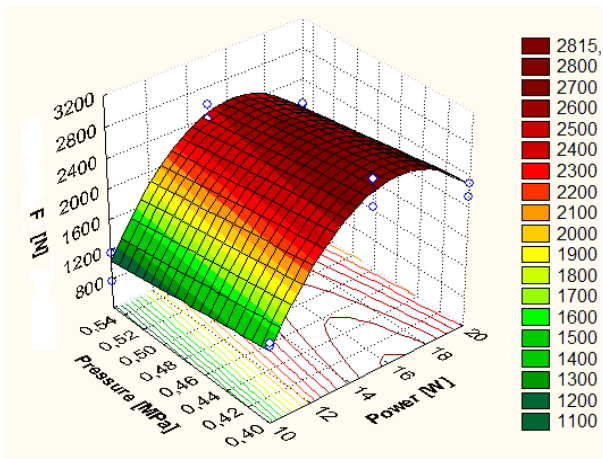
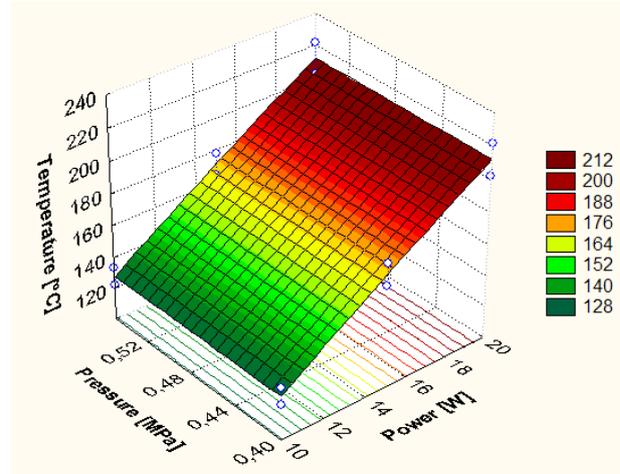

*(a)* *(b)*

*Figure 7 Response surface for the failure force (a) and the simulated temperature at the interface (b) for the exploratory tests*

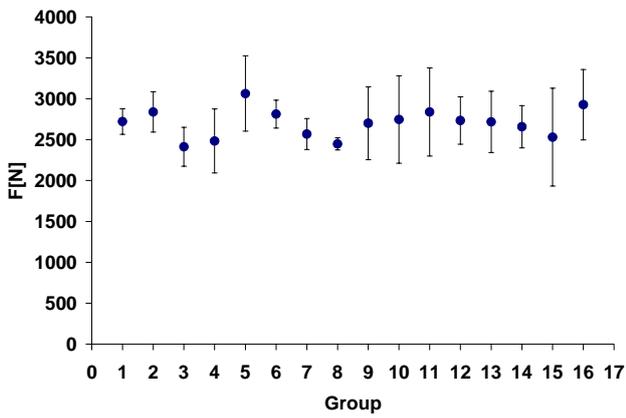
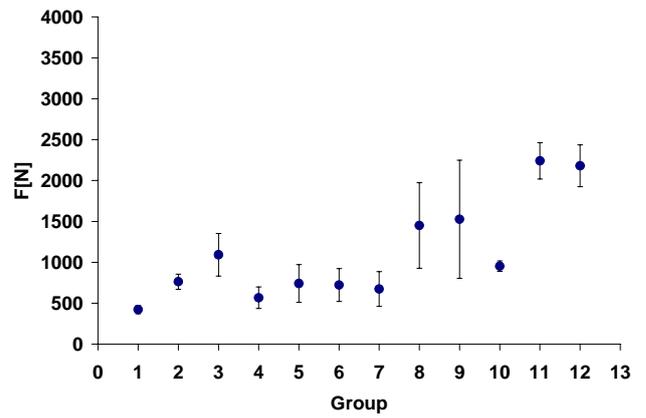

*a) two passages* *b) one passage*

*Figure 8 The influence of the number of laser passages on the failure force*

*Table 3. Study of the influence of laser power and clamping pressure for a welding speed v=0.02 [m/s]*

| Parameters combination | P [W] | $E_{lin}$ [J/mm] | Press. [MPa] | T* [°C] | F [N] | Std. dev. [N] |
|---|---|---|---|---|---|---|
| 1 | 48 | 2.4 | 0.3 | 277 | 418 | 39 |
| 2 | 48 | 2.4 | 0.4 | 277 | 759 | 48 |
| 3 | 48 | 2.4 | 0.5 | 277 | 1034 | 54 |
| 4 | 44 | 2.2 | 0.3 | 257 | 570 | 27 |
| 5 | 44 | 2.2 | 0.4 | 257 | 741 | 63 |
| 6 | 44 | 2.2 | 0.5 | 257 | 718 | 57 |
| 7 | 40 | 2.0 | 0.3 | 239 | 627 | 37 |
| 8 | 40 | 2.0 | 0.4 | 239 | 1341 | 68 |
| 9 | 40 | 2.0 | 0.5 | 239 | 1522 | 62 |
| 10 | 36 | 1.8 | 0.3 | 222 | 974 | 61 |
| 11 | 36 | 1.8 | 0.4 | 222 | 2243 | 44 |
| 12 | 36 | 1.8 | 0.5 | 222 | 2181 | 56 |

\* *simulated temperature at the interface*

*Table 5 Exploratory experimental tests for finding the optimal domain*

| Parameters combination | P [W] | V [m/min] | $E_{lin}$ [J/mm] | Pressure [MPa] | T [°C] | F [N] | Std. dev. [N] |
|---|---|---|---|---|---|---|---|
| 23 | 10 | 0.6 | 1.0 | 0.4 | 122 | 1405 | 57 |
| 24 | 10 | 0.5 | 1.2 | 0.4 | 133 | 1454 | 68 |
| 25 | 10 | 0.6 | 1.0 | 0.55 | 122 | 768 | 52 |
| 26 | 10 | 0.5 | 1.2 | 0.55 | 133 | 1149 | 63 |
| 27 | 15 | 0.6 | 1.5 | 0.4 | 164 | 2911 | 27 |
| 28 | 15 | 0.5 | 1.8 | 0.4 | 178 | 2574 | 22 |
| 29 | 15 | 0.6 | 1.5 | 0.55 | 164 | 2565 | 31 |
| 30 | 15 | 0.5 | 1.8 | 0.55 | 178 | 2381 | 19 |
| 31 | 20 | 0.6 | 2.0 | 0.4 | 202 | 2296 | 36 |
| 32 | 20 | 0.5 | 2.4 | 0.4 | 222 | 2132 | 30 |
| 33 | 20 | 0.6 | 2.0 | 0.55 | 202 | 2049 | 45 |
| 34 | 20 | 0.5 | 2.4 | 0.55 | 222 | 2065 | 32 |